\documentclass[11pt,aps,nofootinbib,superscriptaddress,amssymb]{revtex4}

\usepackage{amsmath,setspace,subfigure,amsfonts,latexsym}
\usepackage{amssymb}
\usepackage{color}
\usepackage{epsfig}
\usepackage{color}
\usepackage{hyperref}
\usepackage[compat=1.1.0]{tikz-feynman}
\tikzfeynmanset{
every edge={very thick},
}

\newcommand{\dis}[1]{\begin{equation}\begin{split}#1\end{split}\end{equation}}
\newcommand{\be}{\begin{equation}}
\newcommand{\ee}{\end{equation}}
\def\bea{\begin{eqnarray}}
\def\eea{\end{eqnarray}}
\newcommand\ba{\begin{eqnarray}}
\newcommand\ea{\end{eqnarray}}

\newcommand{\bfrac}[2]{{\left(\frac{#1}{#2} \right)  }}
\newcommand{\VEV}[1]{\langle #1 \rangle}

\newcommand\gev{\,{\rm GeV}}

\newcommand{\Mp}{M_P}

\newcommand{\mpl}{m_{\rm Pl}}

\newcommand\fa{{f_a}}

\newcommand{\Tmax}{T_{\rm max}}
\newcommand{\Treh}{T_{\rm reh}}

\newcommand\axino{{\tilde{a}}}

\newcommand\gluino{{\tilde{g}}}

\begin{document}

\title{\Large  Axino abundances in high-scale supersymmetry}

\author{Ki-Young Choi}
\affiliation{Department of Physics, BK21 Physics Research Division, Institute of Basic Science, Sungkyunkwan University, Suwon 16419, Korea.}
\email{kiyoungchoi@skku.edu }
\author{Hyun Min Lee}
\affiliation{Department of Physics, Chung-Ang University, Seoul 06974, Korea.}
\affiliation{School of Physics, Korea Institute for Advanced Study, Seoul 02455, Korea.}
\email{hminlee@cau.ac.kr }

\begin{abstract}
\noindent
We consider the thermal production of axino dark matter in high-scale supersymmetry where all the superpartners except the axino are heavier than the maximum and reheating temperatures. In this case, the axinos are produced dominantly in pairs from the scattering of SM particles in thermal plasma in the early Universe after inflation. We find that the thermal averaged scattering cross section for the axino pair production is given by $\VEV{\sigma v} \propto T^4$ in Kim-Shifman-Vainstein-Zakharov (KSVZ) axion model, while it does not depend on the temperature in Dine-Fischler-Srednicki-Zhitnitski (DFSZ) axion model. As a result, the axinos produced during the early matter domination is diluted by the entropy production, so the axino abundance is determined mainly by the reheating temperature, unlike the case with gravitino dark matter.
We show that the axino pair production  in DFSZ model opens up new parameter space for axino dark matter, due to non-decoupled Higgsino interactions at tree level. 
\end{abstract}

\maketitle

\titlepage

\tableofcontents

\newpage

\section{Introduction}
As a solution to the strong CP problem in the Standard Model (SM), the axion has been introduced as a pseudo-Goldstone boson obtained after a spontaneous breakdown of the $U(1)$  Peccei-Quinn (PQ) symmetry. The axino  is the supersymmetric partner of the axion. If R-parity is unbroken and the axino is the lightest supersymmetric particle (LSP), the axino can be a good dark matter (DM) candidate. We note that the recent Planck 2018 result~\cite{Aghanim:2018eyx} determines the DM relic density,
\dis{
 \Omega_{\rm CDM} h^2 = 0.120 \pm 0.001.
 }
 
The effective interaction of the axion at low energies includes the couplings to gluons, suppressed by the PQ scale $\fa$.
Then, the astrophysical and cosmological bounds~\cite{Kim:2008hd} constrain $\fa$ to be
\dis{
5\times 10^{8}\gev \lesssim \fa \lesssim 10^{12}\gev,
}
although the upper bound may be relaxed depending on the energy scale of inflation. Since the axino is the superpartner of the axion, the supersymmetric interaction of axino to gluon and gluino is also suppressed by the same scale $\fa$. Due to such a large suppression of the interactions, the axino freezes out  from the thermal plasma at a high temperature, $T_f\sim 10^{11}\gev$ for $\fa=10^{12}\gev$, so the axino abundance is sensitive to the early Universe dynamics.

One of the important variables in the early Universe is the reheating temperature $\Treh$, which is defined roughly by the onset of the radiation-dominated (RD) era after the early matter domination (eMD). For example,  after a slow-roll inflation ends, the inflaton oscillates around the minimum of the potential and it dominates the energy density of the Universe as a non-relativistic matter. When the Hubble expansion rate $H$ is similar to  the decay rate of the inflaton $\Gamma_\phi$, namely, $H\sim \Gamma_\phi$, the inflaton eventually decays and makes the transition from eMD to RD happen.

When the reheating temperature is sufficiently high such that $\Treh > T_f$, the axinos could be in thermal equilibrium and decoupled while still relativistic. 
In this case, the axino with mass around keV  can play a role of warm dark matter~\cite{Rajagopal:1990yx}. On the other hand, when the reheating temperature is lower than the axino freeze-out temperature, that is, $\Treh < T_f$, the axinos could never be in thermal equilibrium, but instead a small amount of axinos could be produced from thermal particles.
This possibility is called E-WIMP~\cite{Choi:2005vq}, super-WIMP~\cite{Feng:2003xh} or FIMP~\cite{Hall:2009bx} in the literature. In this case, the axino abundance depends on the reheating temperature and/or masses of the heaviest particles, through the scattering and/or decay processes for producing the axinos~\cite{Covi:2001nw}.

Another possibility to produce axinos is through the non-thermal production mechanism. After heavy particles freeze out at high temperature, then they decay later into axinos.
 For the axino with mass around GeV, the non-thermal production can lead to a dominant contribution to the relic density of the axino, that was considered to be a cold  dark matter candidate in~\cite{Covi:1999ty}. 
 Therefore, the axino can take a wide range of masses  depending on the reheating temperature to be dark matter~\cite{Choi:2011yf}.
Even the heavier axino, though it does not saturate the DM relic density, can affect the relic density of the LSP neutralino dark matter~\cite{Choi:2008zq}.

In the literature, however, the single production of the axino from thermal plasma was focused on. In this case, as the thermal averaged scattering cross section for the  axino single production is constant with respect to the temperature,  the axino relic density becomes proportional to the reheating temperature. On the other hand, the axino pair production was neglected, because the corresponding cross section is suppressed by the fourth inverse power of $f_a$, as compared to the second inverse power suppression for the axino single production.

In this article, we consider the axino production in KSVZ and DFSZ axion models with high-scale supersymmetry. In this case, when the maximum and reheating temperatures are smaller than the masses of superparticles other than the axino, the axino single production is exponentially suppressed due to the Boltzmann factors of the equilibrium number densities of superparticles. 
Instead, the axinos can be still produced in pairs due to $R$-parity from the scattering of SM particles in thermal plasma. Therefore, we discuss the effects of the axino pair production in both axion models.

The paper is organized as follows.
We begin with a review on the general thermal and non-thermal production mechanism of the axino in Section \ref{production}.  Then, we present the axino abundances in KSVZ and DFSZ models, in Section \ref{KSVZ} and \ref{DFSZ}, respectively.  The conclusions are drawn in Section \ref{conclusion}.

\section{Thermal and non-thermal production of axinos}
\label{production}

We first review the thermal production of the axinos in the early Universe and define the reheating and maximum  temperatures of the Universe after inflation.
Then, in both KSVZ~\cite{Kim:1979if,Shifman:1979if} and DFSZ~\cite{Dine:1981rt,Zhitnitskii} axion models with low-scale SUSY, we briefly explain the typical mechanisms for the axino single production from the scattering or decay of superparticles in thermal equilibrium.

\subsection{Axino abundances}

The thermal production includes the production from thermal plasmas, through the scattering and decay of  particles in thermal equilibrium. 
Then, we obtain the number density  by solving the Boltzmann equation during the evolution of the Universe:
\dis{
\frac{dn_{\axino}}{dt}+3H n_{\axino} \simeq  \sum_{i,j} \VEV{\sigma (i+j\rightarrow \axino + \cdots)v_{\rm rel}}n_in _j+ \sum_i \VEV{\Gamma(i \rightarrow \axino+ \cdots )}n_i,
 \label{Boltz}
}
where $\VEV{\sigma v}$ and  $\VEV{\Gamma}$  are the thermal averaged scattering cross section and decay rate, respectively, relevant for the axino production, and $n_j$ are the number densities of thermal particles in the relevant processes. Here, we ignored the inverse scattering and inverse decay processes, since they are suppressed due to the small number density of the axino.
Then, the thermal production of axinos depends not only on  the interaction rate of the axinos, $\VEV{\sigma v}$ and  $\VEV{\Gamma}$,  but also the integration range of the Boltzmann equation, which is related to the reheating temperature and/or the maximum  temperature of the Universe as well as the masses of thermal particles.

%Assuming that the axion abundance is determined by thermal production, 
With the final abundance of axinos, we obtain the present relic density as
\dis{
 \Omega_\axino h^2 = \frac{m_\axino n_\axino}{\rho_c/h^2} =0.12 \Big(\frac{m_{\tilde a}}{{\rm GeV}} \Big) \Big(\frac{Y_{\tilde a}}{3.1\times 10^{-9}} \Big),
}
where
\dis{
Y_\axino(T) \equiv \frac{n_\axino}{n_{\rm rad}}, \qquad {\rm and}\qquad  n_{\rm rad} = \frac{\zeta(3)T^3}{\pi^2}.
}

\subsection{Early matter domination and reheating temperature}

The reheating temperature and the maximum temperature of the radiation-dominated Universe depends on the early dynamics before RD.
In many scenarios of the early Universe, the RD is realized due to the decay of heavy particles, which dominate the Universe as non-relativistic matter, through the  early matter domination.  The candidates for heavy particles are inflaton, moduli, curvaton, gravitino, etc.

In the inflationary scenarios, 
when inflation ends, the inflaton field $\phi$ starts oscillating around the minimum of the potential and soon dominates the energy density of the Universe, making the early matter domination possible. Even during eMD, the inflaton field continuously decays and produces light particles, in turn thermalizing rapidly. Then,  the thermal particles produced from the inflaton decay define the temperature, which reaches the maximum temperature $\Tmax$. However, the abundances of thermal particles are at the same time diluted by the entropy production after the inflaton decay.  

Finally, when $H\sim \Gamma_\phi $, where $\Gamma_\phi$ is the decay rate of the inflaton and $H$ is the Hubble expansion rate, most of the inflaton energy is converted into relativistic plasmas, which dominates the energy density and defines the reheating temperature.  
Under the assumption of instantaneous reheating,  the reheating temperature would be then given by
\dis{
\Treh =\bfrac{40}{g\pi^2}^{1/4}\sqrt{\frac{\Gamma_\phi \Mp}{c}},
}
where $\Mp$ is the reduced Planck mass, $g$ is the number of effective degrees of freedom  in the thermal plasma, and $c=1$ for $t_{\rm reh}=\Gamma^{-1}_\phi$ or $c=\frac{2}{3}$ for $\Gamma_\phi=H$. 
However, due to the continuous decay of the inflaton during eMD, the maximum temperature during eMD is higher than the reheating temperature and it is roughly given by $\Tmax\simeq 0.5 \bfrac{m_\phi}{\Gamma_\phi}^{1/4}\Treh$ where $m_\phi$ is the inflaton mass~\cite{Ellis:2015jpg}.

For example, if the inflaton decays by gravitational interactions, then its decay rate is given by
\dis{
\Gamma_\phi\simeq \frac{m_\phi^3}{4\pi \Mp^2},
}
determining the reheating and maximum temperatures by 
\dis{
 \Treh\simeq \bfrac{40}{g\pi^2}^{1/4}    \bfrac{m_\phi^3}{4\pi\Mp}^{1/2},\,\,  {\rm and}\quad \Tmax\simeq 0.5\bfrac{40}{g\pi^2}^{1/4}m_\phi.
 \label{exphi}
}

\subsection{Axino single production: $m_{\rm SUSY} < \Treh <T_f$}

If the reheating temperature is smaller than the freeze-out temperature of the axino, i.e. $\Treh < T_f$, the axinos could not be in thermal equilibrium and  the number density is much suppressed than that of photon. 
However, in low-scale SUSY with $m_{\rm SUSY} < \Treh$, SUSY particles are abundant in thermal plasma, relevant for the axino production from their scattering processes.

When $R$-parity is conserved, the even number  of SUSY particles must participate in the initial and final states for scattering, so we need at least one SUSY particle in thermal equilibrium for the single axino production by scattering processes.  A lot of works on the calculation of the axino abundances in the literature have been focussed on the axino single production~\cite{Covi:1999ty,Covi:2001nw,Choi:2011yf,Brandenburg:2004du,Strumia:2010aa}, because the interactions for the axino pair production are suppressed by one more inverse power of $f_a$.

In KSVZ axion model, the thermal averaged scattering cross section for the axino single production is  dominated by the supersymmetric interactions of QCD anomalies, which is independent of the temperature and given by
\dis{
\langle \sigma v\rangle \simeq \frac{\alpha_s^3}{4\pi^2f_a^2},
}
where $\alpha_s$ is the strong coupling constant.
Thus, the axino abundance is linearly proportional to the reheating temperature as follows~\cite{Covi:2001nw},
\dis{
Y_\axino =\int_{T_0}^{T_{\rm reh}} \frac{\VEV{\sigma v } n_i n_j }{s H T} dT \propto  \VEV{\sigma v }\Mp  \Treh.
}
The approximate result for $Y_\axino$ in KSVZ model~\cite{Brandenburg:2004du} is given by
\dis{
Y_\axino \simeq 2\times 10^{-7} g_s^6\ln\bfrac{1.211}{g_s}\bfrac{10^{11}\gev}{\fa}^2\bfrac{\Treh}{10^4\gev},
}
where $g_s$ is the $SU(3)_C$ gauge coupling with $\alpha_s=g_s^2/(4\pi)$.

In  DFSZ axion model, on the other hand, the axino single production is dominated by the decay of Higgsinos~\cite{Bae:2011jb,Bae:2011iw}, so the axino abundance is determined around the temperature equal to the Higgsino mass to be, 
\dis{
Y_\axino\simeq\int_{T_0}^{T_{\rm reh}} \frac{ \VEV{\Gamma} n_i }{s H T} dT \propto   \frac{\Mp \Gamma}{\mu^2},
}
where $\mu$ is the Higgsino mass and  the decay rate of Higgsinos into Higgs and axino~\cite{Chun:2011zd} is given by
\dis{
\Gamma\simeq \frac{1}{16\pi} \bfrac{\mu}{\fa}^2\mu.
}
The approximate formula for $Y_\axino$ in DFSZ model~\cite{Chun:2011zd,Bae:2011jb} is given by
\dis{
Y_\axino\simeq 8\times 10^{-5}\bfrac{10^{12}\gev}{\fa}^2\bfrac{\mu}{10^6\gev}. \label{DFSZ-hT}
}
When the temperature is below the Higgsino mass, i.e. $\Treh<\mu$, the axino abundance is suppressed exponentially by $\exp(-\mu/\Treh)$.

\section{Axino pair production in high-scale SUSY}

When SUSY particles other than the axino are heavier than the maximum temperature, that is, $\Treh <  \Tmax< m_{\rm SUSY}$, there are no SUSY particles in thermal equilibrium, so  the axino single production is not available. Instead a pair of axinos can be produced from the scattering of a pair of 
Standard Model particles, SM + SM $\rightarrow \axino+\axino$, with heavy SUSY particles exchanged in the $t$-channels. 
In this case, the axino production from the decay of SUSY particles is not available, because SUSY particles are too heavy to keep in thermal equilibrium.

In this section, we discuss new mechanisms for the axino production in both KSVZ and DFSZ models with high-scale SUSY.

\subsection{Axino production in KSVZ model}
\label{KSVZ}

In KSVZ axion models, new heavy quarks carry PQ charges and induce the axion-gluon-gluon anomaly interactions.
At energy scales lower than the heavy quark masses,
the supersymmetric effective interactions for the axion is given by
\dis{
\mathcal{L}= -\frac{\alpha_s}{2\sqrt{2}\pi \fa } \int d^2\theta \, A\rm{Tr} [W_\alpha W^\alpha] + \rm{h.c.},
}
where  $A=(s+ia)/\sqrt{2} +\sqrt{2}\axino \theta +F_A \theta \theta$, is the axion chiral supermultiplet, and $W_\alpha$ is the vector supermultiplet of gluon fields.
Then, the component Lagrangian includes the axino-gluon-gluino interactions as
\dis{
\mathcal{L}_{\rm KSVZ} \supset i\frac{\alpha_s}{16\pi\fa}\overline{\axino}\gamma_5[\gamma^\mu,\gamma^\nu]\tilde{g}^bG_{\mu\nu}^b.  \label{QCD}
}

In high-scale SUSY, when the gluino mass is larger than the reheating temperature, we can integrate out gluinos with the axino-gluon-gluino interactions in Eq.~(\ref{QCD}) and obtain the dimension-7 effective interactions between two axinos and gluons as follows, 
\bea
{\cal L}^{\tilde g}_{\rm eff}= \frac{2\alpha^2_s}{(16\pi f_a)^2 m_{\tilde g}}\, \overline {{\tilde a}} \sigma^{\mu\nu} \sigma^{\rho\sigma} {\tilde a} G^b_{\mu\nu} G^b_{\rho\sigma}\equiv \frac{1}{\Lambda^3_g}\, \overline {{\tilde a}} \sigma^{\mu\nu} \sigma^{\rho\sigma} {\tilde a} G^b_{\mu\nu} G^b_{\rho\sigma}.
\eea 
Then, the scattering processes, $g+g\rightarrow \axino + \axino$, with gluinos in the $t$-channels,  lead to the dominant contributions for the axino production.  The corresponding thermal averaged cross section for $T\ll m_\gluino$ is given by 
\dis{
\VEV{\sigma v}_{gg} = \frac{9\alpha_s^4}{\pi^5 f_a^4 m_\gluino^2}\, T^4.
}
Therefore, the axinos produced during eDM is diluted due to the $T^4$ dependence, so the final axino abundance after reheating is not sensitive to the maximal temperature but it is determined mainly by the reheating temperature.

%%%%%%%%%%%%%%%%%%%%%%%%%%%%%%%%%%%%%%%%%%%%%
\begin{figure*}[!t]
\begin{center}
\begin{tabular}{cc} 
 \includegraphics[width=0.45\textwidth]{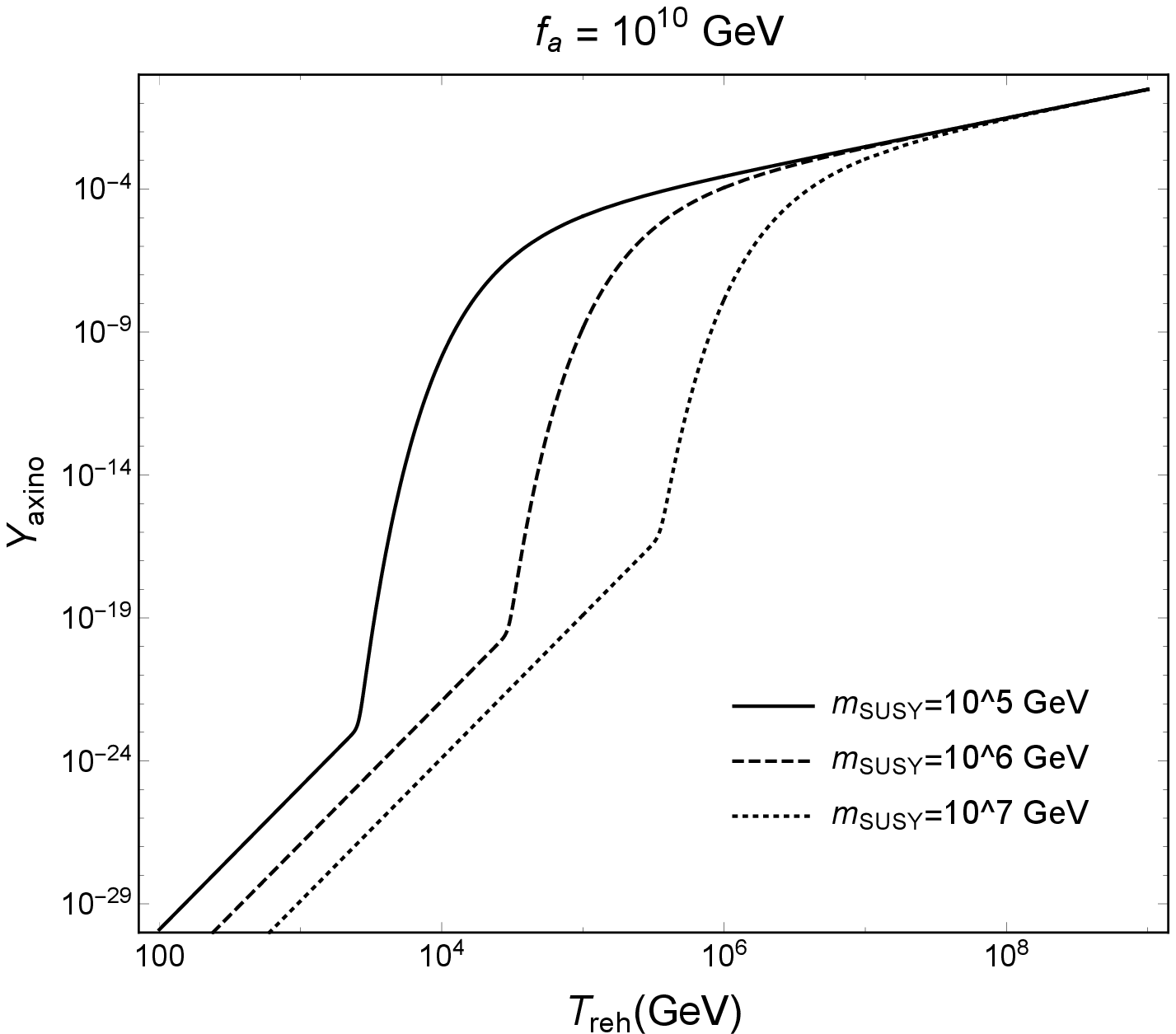}
 &
 \includegraphics[width=0.45\textwidth]{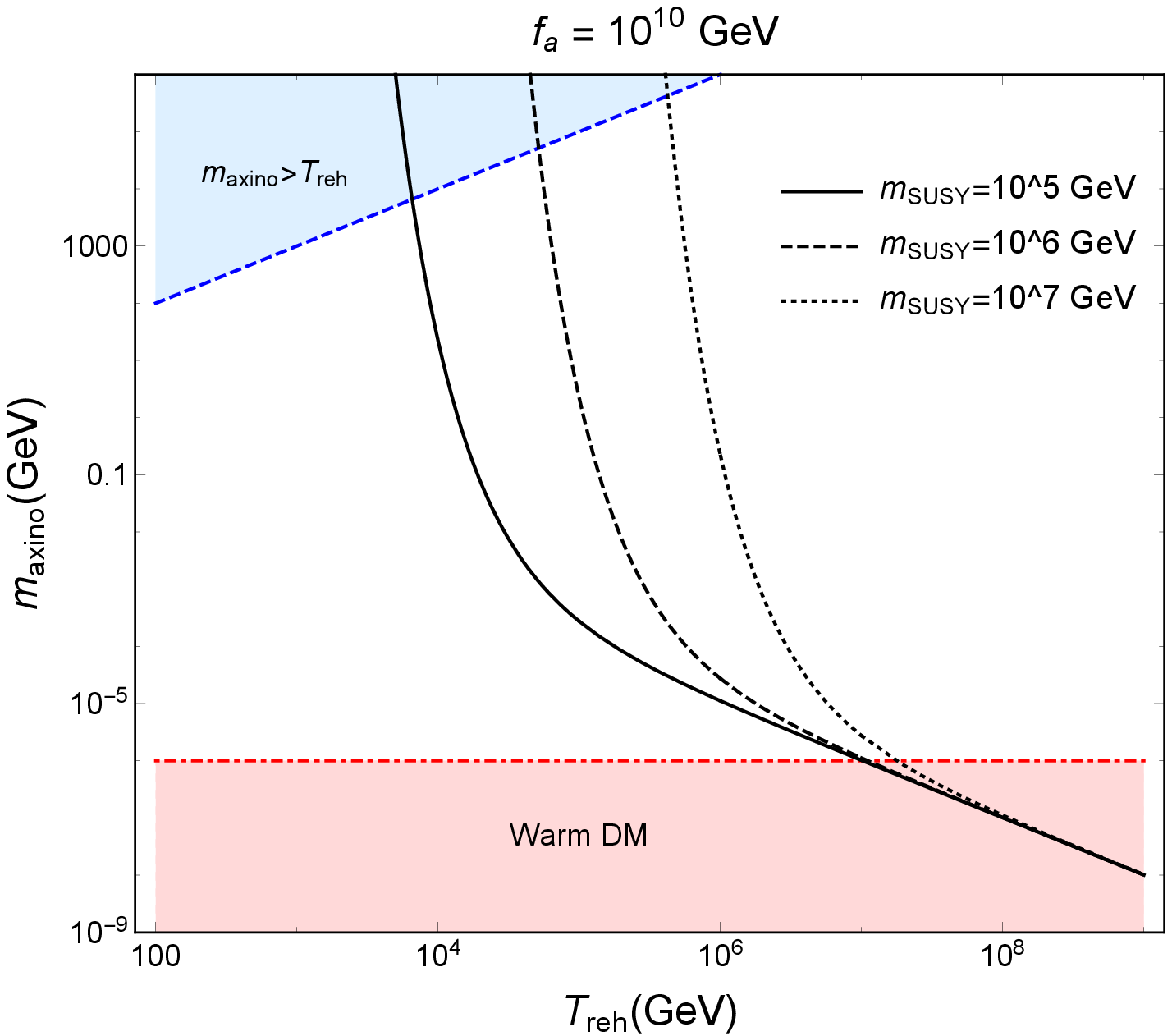}
   \end{tabular}
\end{center}
\caption{Left: Axino abundance as a function of reheating temperature in KSVZ model.  Right: Axino mass to give the correct relic density for dark matter for a given $\Treh$ in the same model. We took $m_{\tilde g}=m_{\rm SUSY}=10^6, 10^6, 10^7 \gev$ in black solid, dashed, and dotted lines, and $\fa=10^{10}\gev$, in both plots. Our result does not apply to the blue region with $m_{\tilde a}>T_{\rm reh}$ and axion becomes a warm dark matter in red region with $m_{\tilde a}<1\,{\rm keV}$. } 
\label{fig:KSVZ}
\end{figure*}
%%%%%%%%%%%%%%%%%%%%%%%%%%%%%%%%%%%%%%%%%%%%%

From the results in Ref.~\cite{Garcia:2017tuj}, in KSVZ model  with high-scale SUSY, the axino abundance is given by 
\bea
Y_{\tilde a}(T) = 4f(4) \Big(1-\frac{\Treh^2}{T^2_{\rm max}}\Big)  Y_{{\tilde a},{\rm inst}} \simeq 2 Y_{{\tilde a},{\rm inst}},
\eea
where $f(4)\simeq 0.5$ numerically corrects the approximate result obtained for instantaneous reheating,
\dis{
Y_{{\tilde a},{\rm inst}}\simeq \Bigg(\frac{90}{g(\Treh)}\Bigg)^{1/2} \Bigg(\frac{g(T)}{g(\Treh)} \Bigg)\frac{36\zeta(3) \alpha_s^4 \Mp \Treh^{5}}{5\pi^8 f_a^4m_{\tilde{g}}^2 },
}
where we get $g(\Treh)=106.75$ in KSVZ model for $m_{t}<T_{\rm reh}<m_{\rm SUSY}$.

In Fig.~\ref{fig:KSVZ}, we depict the axino abundance in KSVZ model as a function of the reheating temperature on left and the parameter space for $m_{\tilde a}$ and $T_{\rm reh}$ on right, satisfying the correct relic density for $m_{\tilde g}=m_{\rm SUSY}=10^5,\, 10^6,\,10^7\,{\rm GeV}$ in black solid, dashed and dotted lines, respectively.  We have chosen the PQ scale to $f_a=10^{10}\,{\rm GeV}$ for both plots.  On the right plot, the blue region is with $m_{\tilde a}>T_{\rm reh}$ for which the axinos could not be produced efficiently from the thermal plasma so our result with axino pair production does not hold. Further, the red region is not favored by large scale structure, because the axino becomes a warm dark matter with $m_{\rm axino}\lesssim 1\,{\rm keV}$.

From the left plot in Fig.~\ref{fig:KSVZ}, we find that the axino abundance shows a slower fall-off at low reheating temperature with $T_{\rm reh}<m_{\rm SUSY}$, due to the axino pair production, unlike the case with axion single production only. 

Comments on the right plot  in Fig.~\ref{fig:KSVZ} are in order. 
First, at high reheating temperature with $T_{\rm reh}>m_{\rm SUSY}$, the axino single production dominates and the relic density does not depend on the other superparticle masses much. 
On the other hand,  for $T_{\rm reh}<m_{\rm SUSY}$, the axino pair production is too small to accommodate a right relic density for $m_{\tilde a}<T_{\rm reh}$ but there is a valid region for the relic density due to the axino single production, although suppressed by the Boltzmann factors of the other superparticles. In this case, a wide range of axino masses can be compatible with the correct relic density, due to the axino single production.

\subsection{Axino production in DFSZ model}
\label{DFSZ}

In DFSZ axion models, the SM fermions carry PQ charges, so the axion-gluon-gluon interactions are generated by the SM quarks at low energy. 
Below the PQ breaking scale, we can write  the effective interactions between axion and Higgs chiral multiplets at tree level   in the following superpotential,
\bea
W_\mu= \mu \, e^{c_H A/f_a} H_u H_d=\Big(1+\frac{c_H}{f_a}\, A+\frac{c^2_H}{2f^2_a}\, A^2+\cdots \Big)\mu H_u H_d
\eea
where $c_H$ is a constant parameter depending on the PQ charge of the Higgs bilinear $H_u H_d$, and the expansion with the axion chiral multiplets is performed up to a few leading terms.
Thus, we obtain the component Lagrangian for the axino-Higgs-Higgsino interactions as
\bea
{\cal L}_{\rm DFSZ}&\supset&- \mu {\tilde H}_u {\tilde H}_d -  \frac{c_H\mu}{f_a}\, {\tilde a}({\tilde H}_u H_d+ H_u {\tilde H}_d)-\frac{c^2_H \mu}{2f_a^2}\, {\tilde a}{\tilde a} H_u H_d +{\rm h.c.}+\cdots,  \label{Higgsino}
\eea
where $H_u=(H^+_u,H^0_u)^T$ and $H_d=(H^0_d,H^-_d)^T$ and the four-component spinor for the axino is understood by ${\tilde a}=({\tilde a}, {\tilde a}^*)^T$. 
We note that  the above effective interactions in DFSZ model are proportional to the Higgsino mass $\mu$, so they show non-decoupled effects in the limit of heavy Higgsinos. But, those effective interactions are suppressed by $\mu/f_a$, thus the effective theory for the axino multiplet is justified.

After integrating out the Higgsinos, we get the following effective interactions between axions and Higgs doublets,
\bea
{\cal L}^{{\tilde H}}_{\rm eff}= \frac{c^2_H\mu}{2f^2_a}\, {\tilde a}{\tilde a} H_u H_d +{\rm h.c.}. \label{DFSZeff}
\eea
Here, we note that it is important to keep the dimension-5 interactions with two axinos in the above expansion, because they are at the same level as for the effective interactions obtained after Higgsinos are integrated out.

Below the electroweak symmetry breaking (EWSB) scale, the axion has a tree-level Yukawa interactions to quark and squark of order of $m_q/\fa$,
and that to Higgs and Higgsino of $\mu/\fa$.  Then, these tree-level interactions generate the QCD as well as electroweak anomaly interactions.
Above the EWSB scale, however, the QCD anomaly disappears, because the SM quarks are massless, while the Higgsinos still contribute $SU(2)_L$ and $U(1)_Y$ anomaly interactions. 

Therefore, when the reheating temperature is larger than the EWSB scale but smaller than superparticle masses, 
a pair of axinos can be produced from thermal plasmas in two ways. One is  the tree level scattering mediated by Higgsinos in the $t$-channels and the other is through the electroweak anomaly interactions of the axino. However, the latter loop-induced interactions are subdominant as compared to the former tree-level interactions so we ignore the loop-induced interactions in the following discussion.

From Eqs.~(\ref{Higgsino}) or (\ref{DFSZeff}), we find that the scattering cross sections relevant for the axino pair production are given by
\dis{
&\sigma (H_u+H_u\rightarrow  \axino + \axino) = \sigma (H_u^*+H_u^*\rightarrow  \axino + \axino)\\
 =&\sigma (H_d+H_d\rightarrow  \axino + \axino) = \sigma (H_d^*+H_d^*\rightarrow  \axino + \axino) \\
   =&4\sigma (H_u+H_d\rightarrow  \axino + \axino) =4\sigma (H_u^*+H_d^*\rightarrow  \axino + \axino) \\
  =&2\sigma (H_p+H_m\rightarrow  \axino + \axino) =\frac{\mu^2c_H^4}{8\pi f_a^4}.
}
Considering the above results, we obtain the total  cross section for the axino pair production as
\dis{
\VEV{\sigma_{\rm tot} v}=  \Big(4+\frac12 +2\times \frac14\Big)\cdot \frac{\mu^2c_H^4}{8\pi f_a^4}= \frac{5\mu^2c_H^4}{8\pi f_a^4}.
}

Therefore, from the results in Ref.~\cite{Garcia:2017tuj}, in DFSZ model with high-scale SUSY, the axino abundance  is given by
\bea
Y_{\tilde a}(T_{\rm reh})  =\frac{4}{15}f(0) \Big(1-\frac{\Treh^6}{T^6_{\rm max}}\Big)  Y_{{\tilde a},{\rm inst}}  \simeq 0.9 Y_{{\tilde a},{\rm inst}} 
\eea
where $f(0)\simeq 3.4$ numerically corrects the approximate result, which is obtained for instantaneous reheating as follows,
\dis{
Y_{{\tilde a},{\rm inst}}\simeq \Bigg(\frac{90}{g(\Treh)}\Bigg)^{1/2} \Bigg(\frac{g(T)}{g(\Treh)} \Bigg)\frac{5\zeta(3)c_H^4 \mu^2 \Mp \Treh}{2\pi^4 f_a^4 }\,.
}
Here, we note $g(\Treh)=110.75$ in DFSZ model with $m_t<T_{\rm reh}<m_{\rm SUSY}$.

%%%%%%%%%%%%%%%%%%%%%%%%%%%%%%%%%%%%%%%%%%%%%
\begin{figure*}[!t]
\begin{center}
\begin{tabular}{cc} 
 \includegraphics[width=0.45\textwidth]{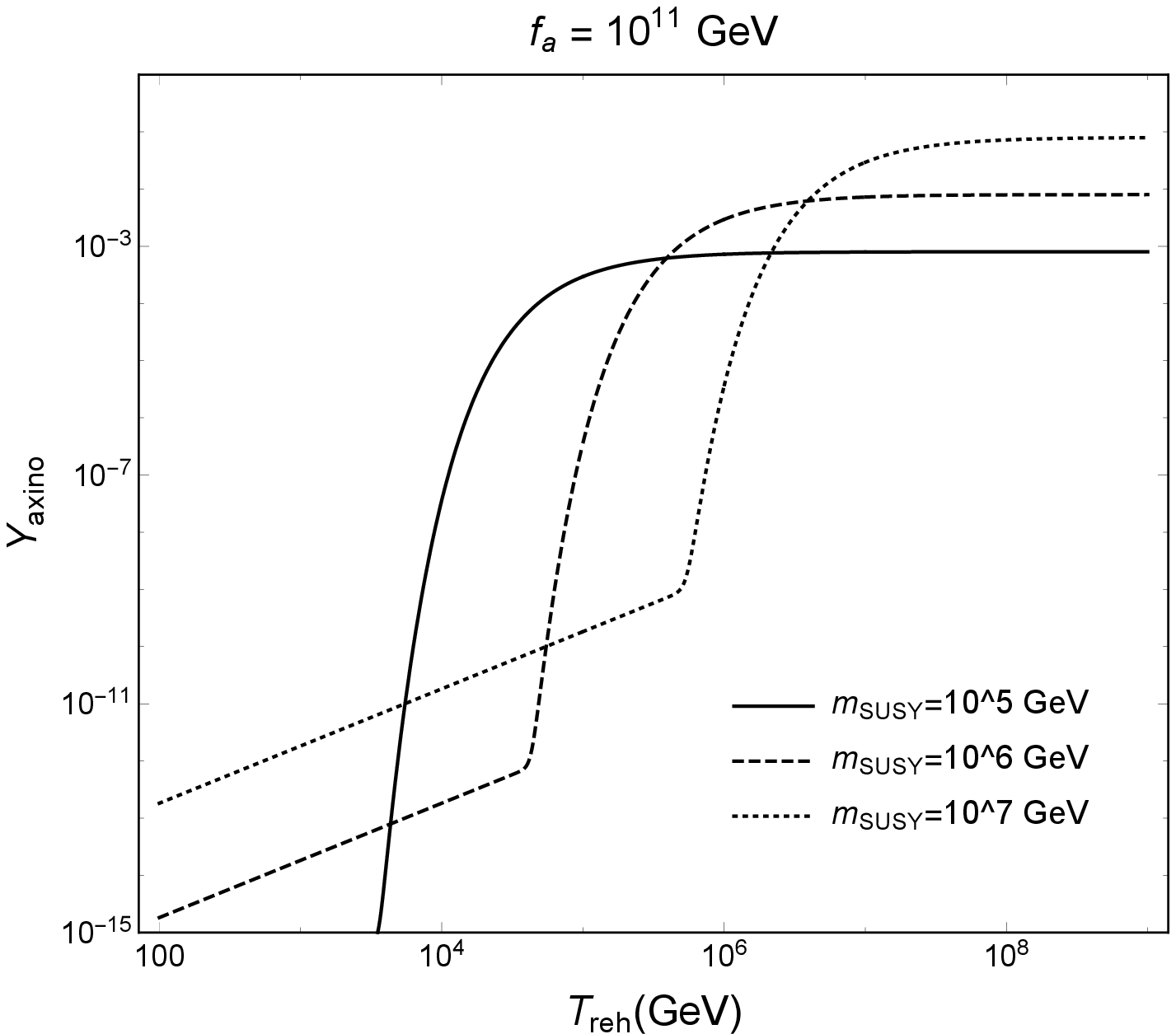}
 &
 \includegraphics[width=0.45\textwidth]{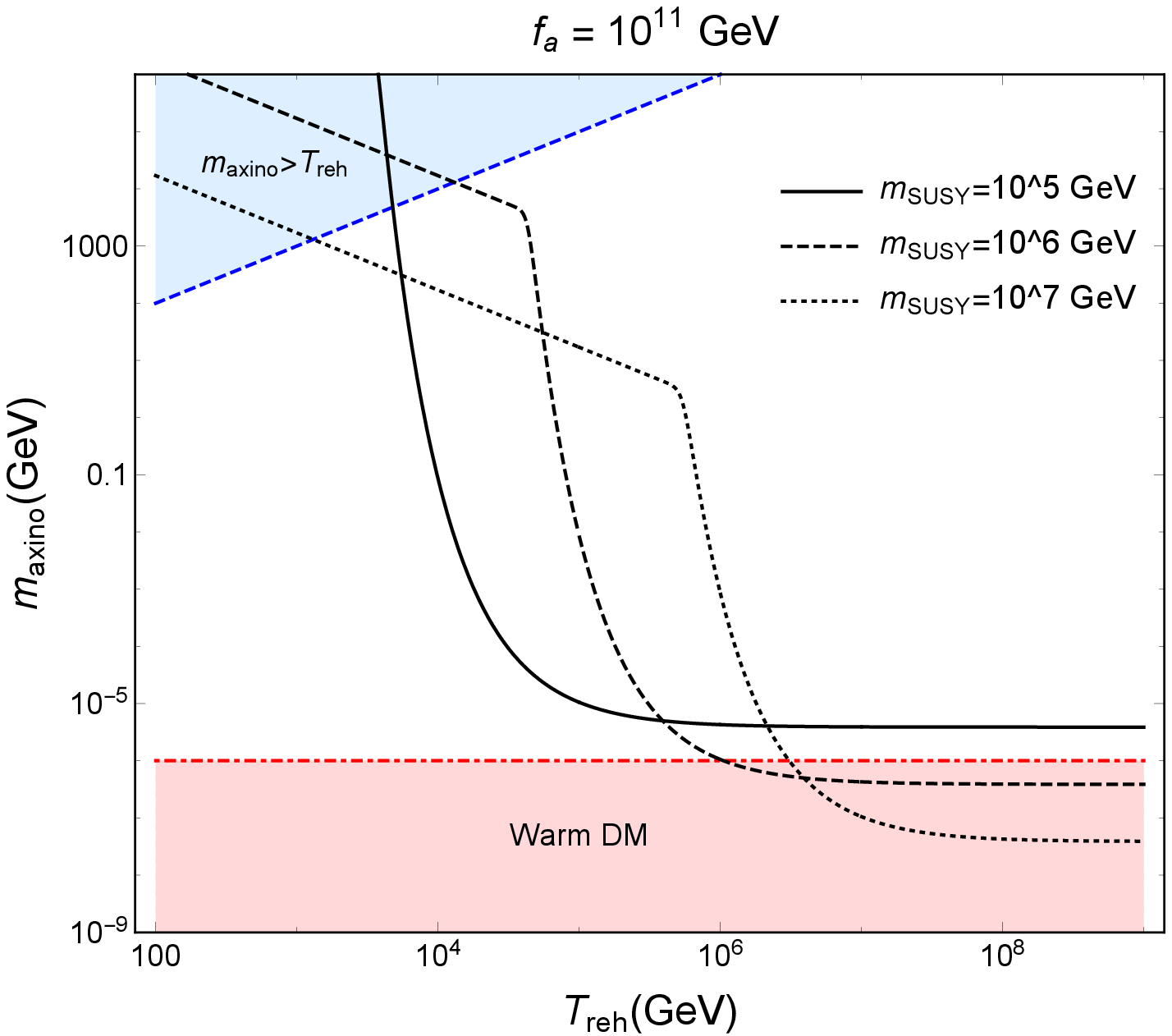}
   \end{tabular}
\end{center}
\caption{Left: Axino abundance as a function of reheating temperature in DFSZ model.  Right: Axino mass to give the correct relic density for dark matter for a given $\Treh$ in the same model.
We took $\mu=m_{\rm SUSY}=10^6, 10^6, 10^7 \gev$ in black solid, dashed, and dotted lines, and $\fa=10^{11}\gev$, in both plots. Our result does not apply to the blue region with $m_{\tilde a}>T_{\rm reh}$ and axion becomes a warm dark matter in red region with $m_{\tilde a}<1\,{\rm keV}$. } 
\label{fig:DFSZ}
\end{figure*}
%%%%%%%%%%%%%%%%%%%%%%%%%%%%%%%%%%%%%%%%%%%%%

In Fig.~\ref{fig:DFSZ}, we show the axino abundance in DFSZ model as a function of the reheating temperature on left and the parameter space for $m_{\tilde a}$ and $T_{\rm reh}$ on right, satisfying the correct relic density for $\mu=m_{\rm SUSY}=10^5,\, 10^6,\,10^7\,{\rm GeV}$ in black solid, dashed and dotted lines, respectively.  We have chosen the PQ scale to $f_a=10^{11}\,{\rm GeV}$ for both plots.  On the right plot, the blue region with $m_{\tilde a}>T_{\rm reh}$ and the red region with $m_{\rm axino}\lesssim 1\,{\rm keV}$ are out of our consideration, due to the invalidity of our calculation of the axino abundance and the problem of large scale structure, respectively, as in Fig.~\ref{fig:KSVZ}

From the left plot in Fig.~\ref{fig:DFSZ}, we find that the axino abundance remains sizable at low reheating temperature with $T_{\rm reh}<m_{\rm SUSY}$, due to the axino pair production, and it saturates to the fixed abundance at high temperature as soon as the axino single production opens up \cite{Chun:2011zd,Bae:2011jb}. 

We also remark some comments on the right plot  in Fig.~\ref{fig:DFSZ}.
First, for $T_{\rm reh}>m_{\rm SUSY}=\mu$, the axino abundance is saturated to a fixed value  \cite{Chun:2011zd,Bae:2011jb}, being proportional to the $\mu$ parameter as in eq.~(\ref{DFSZ-hT}), so the axino mass is accordingly fixed for a correct relic density. In this case, a relatively light axino with $m_{\tilde a}\lesssim 10\,{\rm keV}$ is needed for a correct relic density with $\mu=m_{\rm SUSY}>10^5\,{\rm GeV}$.
On the other hand, for $T_{\rm reh}<m_{\rm SUSY}=\mu$, new parameter space with heavy axino masses opens up at low reheating temperature due to the axino pair production. For instance, in some benchmark points with the Higgsino mass between $\mu=10^5\,{\rm GeV}$ and $10^7\,{\rm GeV}$, the axino masses in the range of $1\,{\rm GeV}\lesssim m_{\tilde a}\lesssim 10\,{\rm TeV}$ are newly allowed. For heavier Higgsino masses, similar results can be obtained but with lighter axino masses.

\section{Conclusions}
\label{conclusion}

We have computed the abundances of axino dark matter in supersymmetric KSVZ and DFSZ models where all the superpartners except the axino are heavier than the maximum and reheating temperatures. 
In this case, the axino single production from the decays and scattering of other superparticles is suppressed by the Boltzmann factor in the presence of $R$-parity, but rather the axino pair production from the scattering of SM particles becomes dominant. 

As a result, we showed that the axino abundances are determined mainly by the reheating temperature, because the thermal averaged scattering cross section for the axino pair production is less sensitive to the temperature than the case for gravitino dark matter, namely, $\langle\sigma v\rangle\propto T^4$ and $T^0$, in KSVZ and DFSZ models, respectively. In KSVZ model, in the new region with $m_{\tilde a}<T_{\rm reh}<m_{\rm SUSY}$ that we considered, we found that the axino pair production is too small to allow for a correct relic density whereas the axino single production can saturate the relic density with a relatively heavy axino mass. On the other hand,  in DFSZ model, we found that new parameter space opens up for a correct relic density at $T_{\rm reh}<m_{\rm SUSY}$, because the axino pair production is efficient due to the tree-level Higgsino interactions.  We showed some benchmark points where axino dark matter with mass $1\,{\rm GeV}\lesssim m_{\tilde a}\lesssim 10\,{\rm TeV}$ is newly allowed, depending on the Higgsino mass between $ 10^5\,{\rm GeV}$ and $10^7\,{\rm GeV}$ for $f_a=10^{11}\,{\rm GeV}$.

\section*{Acknowledgments}

The work of K.-Y.C. is supported in part by the National Research Foundation of Korea(NRF) grant funded by the Korea government(MEST) (NRF-2016R1A2B4012302).
The work of HML is supported in part by Basic Science Research Program through the National Research Foundation of Korea (NRF) funded by the Ministry of Education, Science and Technology (NRF-2016R1A2B4008759 and NRF-2018R1A4A1025334).

\def\theequation{A.\arabic{equation}}

\setcounter{equation}{0}

\vskip0.8cm
\noindent
{\Large \bf Appendix A: Thermal averaged scattering cross sections} 
\vskip0.4cm
\noindent

The thermal averaged scattering cross section~\cite{Gondolo:1990dk} is given by 
\dis{
\langle \sigma v\rangle= \frac{\int d^3p_1 d^3 p_2\, e^{-(E_1+E_2)/T}\, \sigma v}{\int d^3p_1 d^3 p_2\,e^{-(E_1+E_2)/T }}. \label{averaged}
}
For $\sigma = s^n$, we then obtain the numerator,
\dis{
\int d^3p_1 d^3 p_2\, e^{-(E_1+E_2)/T}\, \sigma v &= 2\pi^2 T \int ds \sigma (s-4m^2) \sqrt{s}K_1(\sqrt{s}/T)\\
&= \pi^2 2^{2n+5} T^{2n+6} \Gamma(n+2)\Gamma(n+3),
}
where we ignored particle masses in the second line.
Taking $n=0$ in the above formula, the denominator in eq.~(\ref{averaged})  is given by $64\pi^2 T^6$.
Therefore, we get the general formula for the thermal averaged cross section,
\dis{
\langle s^n v\rangle= 2^{2n-1}T^{2n}\Gamma(n+2)\Gamma(n+3).
}

 %%%%%%%%%%%%%%%%%%%%%%%%%%%%%%%%%%%%%%%%%%%%%%%%%%%%%%%%%%%%

%%%\section*{References}
%\bibliography{mybibfile}

\begin{thebibliography}{999}

  
%\cite{Aghanim:2018eyx}
\bibitem{Aghanim:2018eyx}
  N.~Aghanim {\it et al.} [Planck Collaboration],
  %``Planck 2018 results. VI. Cosmological parameters,''
  arXiv:1807.06209 [astro-ph.CO].
  %%CITATION = ARXIV:1807.06209;%%
  
    %\cite{Kim:2008hd}
\bibitem{Kim:2008hd}
  J.~E.~Kim and G.~Carosi,
  %``Axions and the Strong CP Problem,''
  Rev.\ Mod.\ Phys.\  {\bf 82} (2010) 557.
%  doi:10.1103/RevModPhys.82.557
%  [arXiv:0807.3125 [hep-ph]].
  %%CITATION = doi:10.1103/RevModPhys.82.557;%% 
  
  %\cite{Rajagopal:1990yx}
\bibitem{Rajagopal:1990yx}
  K.~Rajagopal, M.~S.~Turner and F.~Wilczek,
  %``Cosmological implications of axinos,''
  Nucl.\ Phys.\ B {\bf 358} (1991) 447.
  doi:10.1016/0550-3213(91)90355-2
  %%CITATION = doi:10.1016/0550-3213(91)90355-2;%%
  
  
  %\cite{Choi:2005vq}
\bibitem{Choi:2005vq}
  K.~Y.~Choi and L.~Roszkowski,
  %``E-WIMPs,''
  AIP Conf.\ Proc.\  {\bf 805} (2006) 30
  doi:10.1063/1.2149672
  [hep-ph/0511003].
  %%CITATION = doi:10.1063/1.2149672;%%
  
  %\cite{Feng:2003xh}
\bibitem{Feng:2003xh}
  J.~L.~Feng, A.~Rajaraman and F.~Takayama,
  %``Superweakly interacting massive particles,''
  Phys.\ Rev.\ Lett.\  {\bf 91} (2003) 011302
  doi:10.1103/PhysRevLett.91.011302
  [hep-ph/0302215].
  %%CITATION = doi:10.1103/PhysRevLett.91.011302;%%
  
  %\cite{Hall:2009bx}
\bibitem{Hall:2009bx}
  L.~J.~Hall, K.~Jedamzik, J.~March-Russell and S.~M.~West,
  %``Freeze-In Production of FIMP Dark Matter,''
  JHEP {\bf 1003} (2010) 080
  doi:10.1007/JHEP03(2010)080
  [arXiv:0911.1120 [hep-ph]].
  %%CITATION = doi:10.1007/JHEP03(2010)080;%%
  

  %\cite{Covi:2001nw}
\bibitem{Covi:2001nw}
  L.~Covi, H.~B.~Kim, J.~E.~Kim and L.~Roszkowski,
  %``Axinos as dark matter,''
  JHEP {\bf 0105} (2001) 033.
%  doi:10.1088/1126-6708/2001/05/033
%  [hep-ph/0101009].
  %%CITATION = doi:10.1088/1126-6708/2001/05/033;%%
  
  
%\cite{Covi:1999ty}
\bibitem{Covi:1999ty}
  L.~Covi, J.~E.~Kim and L.~Roszkowski,
  %``Axinos as cold dark matter,''
  Phys.\ Rev.\ Lett.\  {\bf 82} (1999) 4180.
%  doi:10.1103/PhysRevLett.82.4180
%  [hep-ph/9905212].
  %%CITATION = doi:10.1103/PhysRevLett.82.4180;%%
  
  

%\cite{Choi:2011yf}
\bibitem{Choi:2011yf}
  K.~Y.~Choi, L.~Covi, J.~E.~Kim and L.~Roszkowski,
  %``Axino Cold Dark Matter Revisited,''
  JHEP {\bf 1204} (2012) 106.
%  doi:10.1007/JHEP04(2012)106
%  [arXiv:1108.2282 [hep-ph]].
  %%CITATION = doi:10.1007/JHEP04(2012)106;%%


%\cite{Choi:2008zq}
\bibitem{Choi:2008zq}
  K.~Y.~Choi, J.~E.~Kim, H.~M.~Lee and O.~Seto,
  %``Neutralino dark matter from heavy axino decay,''
  Phys.\ Rev.\ D {\bf 77} (2008) 123501.
%  doi:10.1103/PhysRevD.77.123501
%  [arXiv:0801.0491 [hep-ph]].
  %%CITATION = doi:10.1103/PhysRevD.77.123501;%%

  
%\cite{Kim:1979if}
\bibitem{Kim:1979if}
  J.~E.~Kim,
  %``Weak Interaction Singlet and Strong CP Invariance,''
  Phys.\ Rev.\ Lett.\  {\bf 43} (1979) 103.
  doi:10.1103/PhysRevLett.43.103
  %%CITATION = doi:10.1103/PhysRevLett.43.103;%%
  
  %\cite{Shifman:1979if}
\bibitem{Shifman:1979if}
  M.~A.~Shifman, A.~I.~Vainshtein and V.~I.~Zakharov,
  %``Can Confinement Ensure Natural CP Invariance of Strong Interactions?,''
  Nucl.\ Phys.\ B {\bf 166} (1980) 493.
  doi:10.1016/0550-3213(80)90209-6
  %%CITATION = doi:10.1016/0550-3213(80)90209-6;%%
  
  %\cite{Dine:1981rt}
\bibitem{Dine:1981rt}
  M.~Dine, W.~Fischler and M.~Srednicki,
  %``A Simple Solution to the Strong CP Problem with a Harmless Axion,''
  Phys.\ Lett.\  {\bf 104B} (1981) 199.
  doi:10.1016/0370-2693(81)90590-6
  %%CITATION = doi:10.1016/0370-2693(81)90590-6;%%
  
  \bibitem{Zhitnitskii}
   A. P. Zhitnitskii, Sov. J. Nucl. Phys. {\bf 31}, 260 (1980).


%\cite{Ellis:2015jpg}
\bibitem{Ellis:2015jpg}
  J.~Ellis, M.~A.~G.~Garcia, D.~V.~Nanopoulos, K.~A.~Olive and M.~Peloso,
  %``Post-Inflationary Gravitino Production Revisited,''
  JCAP {\bf 1603} (2016) no.03,  008
  doi:10.1088/1475-7516/2016/03/008
  [arXiv:1512.05701 [astro-ph.CO]].
  %%CITATION = doi:10.1088/1475-7516/2016/03/008;%%
  
%\cite{Brandenburg:2004du}
\bibitem{Brandenburg:2004du}
  A.~Brandenburg and F.~D.~Steffen,
  %``Axino dark matter from thermal production,''
  JCAP {\bf 0408} (2004) 008
  doi:10.1088/1475-7516/2004/08/008
  [hep-ph/0405158].
  %%CITATION = doi:10.1088/1475-7516/2004/08/008;%%
  
  %\cite{Strumia:2010aa}
\bibitem{Strumia:2010aa}
  A.~Strumia,
  %``Thermal production of axino Dark Matter,''
  JHEP {\bf 1006} (2010) 036
  doi:10.1007/JHEP06(2010)036
  [arXiv:1003.5847 [hep-ph]].
  %%CITATION = doi:10.1007/JHEP06(2010)036;%%

  


%\cite{Bae:2011jb}
\bibitem{Bae:2011jb}
  K.~J.~Bae, K.~Choi and S.~H.~Im,
  %``Effective Interactions of Axion Supermultiplet and Thermal Production of Axino Dark Matter,''
  JHEP {\bf 1108} (2011) 065.
%  doi:10.1007/JHEP08(2011)065
%  [arXiv:1106.2452 [hep-ph]].
  %%CITATION = doi:10.1007/JHEP08(2011)065;%%


  %\cite{Bae:2011iw}
\bibitem{Bae:2011iw}
  K.~J.~Bae, E.~J.~Chun and S.~H.~Im,
  %``Cosmology of the DFSZ axino,''
  JCAP {\bf 1203} (2012) 013
  doi:10.1088/1475-7516/2012/03/013
  [arXiv:1111.5962 [hep-ph]].
  %%CITATION = doi:10.1088/1475-7516/2012/03/013;%%



  %\cite{Garcia:2017tuj}
\bibitem{Garcia:2017tuj}
  M.~A.~G.~Garcia, Y.~Mambrini, K.~A.~Olive and M.~Peloso,
  %``Enhancement of the Dark Matter Abundance Before Reheating: Applications to Gravitino Dark Matter,''
  Phys.\ Rev.\ D {\bf 96} (2017) no.10,  103510
  doi:10.1103/PhysRevD.96.103510
  [arXiv:1709.01549 [hep-ph]].
  %%CITATION = doi:10.1103/PhysRevD.96.103510;%%


  %\cite{Chun:2011zd}
\bibitem{Chun:2011zd}
  E.~J.~Chun,
  %``Dark matter in the Kim-Nilles mechanism,''
  Phys.\ Rev.\ D {\bf 84} (2011) 043509
  doi:10.1103/PhysRevD.84.043509
  [arXiv:1104.2219 [hep-ph]].
  %%CITATION = doi:10.1103/PhysRevD.84.043509;%%
  
%\cite{Gondolo:1990dk}
\bibitem{Gondolo:1990dk}
  P.~Gondolo and G.~Gelmini,
  %``Cosmic abundances of stable particles: Improved analysis,''
  Nucl.\ Phys.\ B {\bf 360} (1991) 145.
  doi:10.1016/0550-3213(91)90438-4
  %%CITATION = doi:10.1016/0550-3213(91)90438-4;%%
  %922 citations counted in INSPIRE as of 18 Sep 2018  
  
      
\end{thebibliography}
%%%%%%%%%%%%%%%%%%%%%%%%%%%%%%%%%%%%%%

\def\prp#1#2#3{Phys.\ Rep.\ {\bf #1} #2 (#3)}
\def\rmp#1#2#3{Rev. Mod. Phys.\ {\bf #1}  #2 (#3)}
\def\anrnp#1#2#3{Annu. Rev. Nucl. Part. Sci.\ {\bf #1} #2 (#3)}
\def\npb#1#2#3{Nucl.\ Phys.\ {\bf B#1}  #2 (#3)}
\def\plb#1#2#3{Phys.\ Lett.\ {\bf B#1}  #2 (#3)}
\def\prd#1#2#3{Phys.\ Rev.\ {\bf D#1}, #2  (#3)}
\def\prl#1#2#3{Phys.\ Rev.\ Lett.\ {\bf #1}  #2 (#3)}
\def\jhep#1#2#3{JHEP\ {\bf #1}  #2 (#3)}
\def\jcap#1#2#3{JCAP\ {\bf #1}  #2 (#3)}
\def\zp#1#2#3{Z.\ Phys.\ {\bf #1}  #2 (#3)}
\def\epjc#1#2#3{Euro. Phys. J.\ {\bf #1}  #2 (#3)}
\def\ijmp#1#2#3{Int.\ J.\ Mod.\ Phys.\ {\bf #1}  #2 (#3)}
\def\mpl#1#2#3{Mod.\ Phys.\ Lett.\ {\bf #1}  #2 (#3)}
\def\apj#1#2#3{Astrophys.\ J.\ {\bf #1}  #2 (#3)}
\def\nat#1#2#3{Nature\ {\bf #1}  #2 (#3)}
\def\sjnp#1#2#3{Sov.\ J.\ Nucl.\ Phys.\ {\bf #1}  #2 (#3)}
\def\apj#1#2#3{Astrophys.\ J.\ {\bf #1}  #2 (#3)}
\def\ijmp#1#2#3{Int.\ J.\ Mod.\ Phys.\ {\bf #1}  #2 (#3)}
\def\apph#1#2#3{Astropart.\ Phys.\ {\bf B#1}, #2 (#3) }
\def\mnras#1#2#3{Mon.\ Not.\ R.\ Astron.\ Soc.\ {\bf #1}  #2 (#3)}
\def\nat#1#2#3{Nature (London)\ {\bf #1}  #2 (#3)}
\def\apjs#1#2#3{Astrophys.\ J.\ Supp.\ {\bf #1}  #2 (#3)}
\def\aipcp#1#2#3{AIP Conf.\ Proc.\ {\bf #1}  #2 (#3)}
\def\njp#1#2#3{New\ J.\ Phys.\ {\bf #1} (#3) #2}

\end{document}